\documentclass[twocolumn, showpacs]{revtex4}
\pdfoutput=1
\usepackage[pdftex,dvipsnames]{xcolor}
\usepackage{xargs}
\usepackage{graphicx}
\usepackage{hyperref}
\usepackage{epstopdf}
\usepackage{color}
\usepackage{float}
\usepackage{amsmath}
\usepackage{amsfonts}
\usepackage[utf8]{inputenc}
\usepackage{mathrsfs}
\usepackage{enumerate}

\begin{document}
\title{Energy degeneracies from Broad Histogram Method and Wang-Landau Sampling}

\author{A. P. Lima$^{1,2}$}
\thanks{E-mail: alelima@if.uff.br}

\author{P. M. C. de Oliveira$^{1,3,4}$}

\author{Daniel Girardi$^{5}$}

\affiliation{$^1$Instituto de F\'isica, UFF, Niter\'oi, Brazil\\
$^2$Centro Federal de Educa\c c\~ao T\'ecnica e Tecnol\'ogica Celso Suckow da Fonseca, CEFET-RJ, Itagua\'i, Brazil\\
$^3$Instituto Mercosul de Estudos Avan\c cados, Universidade Federal da Integra\c c\~ao Latino Americana, Foz do Igua\c cu, Brazil\\
$^4$Instituto Nacional de Ci\^encia e Tecnologia, Sistemas Complexos, Rio de Janeiro, Brazil\\
$^5$Departamento de Ci\^encias Exatas e Licenciaturas,  Centro de Blumenau -  Universidade Federal de Santa Catarina, SC, Brazil
}

\begin{abstract}
In this work, we present a comparative study of the accuracy provided by the Wang-Landau sampling and the Broad Histogram method to estimate de density of states of the two dimensional Ising ferromagnet. The microcanonical averages used to describe the thermodynamic behaviour and to use the Broad Histogram method were obtained using the single spin-flip Wang-Landau sampling, attempting to convergence issues and accuracy improvements. We compare the results provided by both techniques with the exact ones for thermodynamic properties and critical exponents. Our results, within the Wang-Landau sampling, reveal that the Broad Histogram approach provides a better description of the density of states for all cases analysed.
\end{abstract}

%



\maketitle

\section{Introduction}

For physical systems subjected to the constrain of fixed number of particles, volume and temperature (canonical equilibrium), the thermal average of some macroscopic quantity $Q$ for a temperature $T$ is given by:
\begin{equation}
\left<Q\right>_T = \frac{\sum_{\lbrace S\rbrace}Q(S)\,\exp[-E(S)/T]}{\sum_{\lbrace S\rbrace}\exp[-E(S)/T]}
\label{mediacanonica},
\end{equation}
taking $k_B=1$. This sum runs over all microstates available for the system, each of which with an energy $E(S)$ associated with the macroscopic quantity $Q(S)$. The aim of the Monte Carlo (MC) simulations used in statistical physics is to provide an approximation to this sum. The Metropolis algorithm \cite{Metropolis1953} has been widely applied to build a representative subset of $\lbrace S\rbrace$. 
Some technical details (see \cite{BinderHeerman, landau2009guide, newman1999monte} for a general reference) have to be deployed to control the quality of this approximation for each temperature. Reweighting methods, as introduced by Salsburg in Ref.~\cite{salsburg1959} and made popular in \cite{PhysRevLett.61.2635, PhysRevLett.63.1195}, can be used to get $\left<Q\right>_T$ for a range $\Delta T$ around $T$. A drawback of this approach is that $\left<Q\right>_T$ is estimated by an arithmetical average, so the configurations sampled should be uncorrelated. This requires an extra computational effort \cite{PhysRevLett.58.86}, due to critical slowing down \cite{WOLFF199093} in continuous phase transitions, where statistical correlations comes into play.


To overcome such obstacles, one can notice that the summations in Eq.~(\ref{mediacanonica}) can be rewritten by grouping microstates $S_i$, which possess the same energy $E(S_i) = E$. Then, factoring each possible value within the system's energy spectrum $\lbrace\mathbf{E}\rbrace$, each one corresponding to a total of $g(E)$ degenerate microstates (i.e., the density of states), we can rewrite Eq.~(\ref{mediacanonica}) as
\begin{equation}
\left<Q\right>_T = \frac{\sum_{\lbrace\mathbf{E}\rbrace} g(E)\left<Q(E)\right>\exp(-E/T)}{\sum_{\lbrace\mathbf{E}\rbrace}g(E)\exp(-E/T)},
\label{mediacan2}
\end{equation}
where the microcanonical average  
\begin{equation}
\left<Q(E)\right> = \frac{\sum_{S(E)} Q(S)}{g(E)}
\label{mediamicro}
\end{equation}
is now restricted to microstates with the same energy $E$, uniformly weighted. Despite the Eq.~(\ref{mediacan2}) be equivalent to (\ref{mediacanonica}), it is conceptually different. Written this way, the canonical averages are obtained through $g(E)$ and $\left<Q(E)\right>$. Both are intrinsic to the physical system itself and do not depend on any imposed constraint or if it is in equilibrium. The density of states tells us how the energy is distributed among the microstates of the system, independent of interactions with the surroundings.

If one wants to use Eq.~(\ref{mediacan2}) to describe the thermal properties of the system, a dynamics that samples all energy levels is necessary. Some simulation algorithms, like in Refs. \cite{PMCO_microcan, Wang2000147}, have been proposed to do so. But the Wang-Landau sampling (WLS) \cite{PhysRevLett.86.2050} has been widely applied to many systems and became a simple and accurate Monte Carlo method. A dynamics-independent method to evaluate $g(E)$ became known as the Broad Histogram Method (BHM) \cite{BHM_BJP}, which uses the macroscopic quantities $N_{up}$ and $N_{dn}$ to recursively obtain the density of states.

In this paper we compare the densities of states obtained with the BHM and WLS for the two dimensional Ising ferromagnet, in order to stablish the accuracy obtained by each method. In Sec. \textbf{II} we describe these two methods to evaluate the density of states and the convergence analysis applied in the use of the Wang-Landau sampling. Sec. \textbf{III} presents a comparative accuracy analysis for the temperature dependent thermodynamic behaviour and characterization of continuous phase transition. Conclusions and final considerations are presented in Sec. \textbf{IV}

\section{Methodology}

To obtain an accurate estimation of the density of states it is necessary to sample all energy levels (not only around $\langle E\rangle_T$, as done by importance sampling). One solution is to perform an unbiased random walk (RW) in the energy axis. To do it, one needs to stablish a stochastic dynamics to change the microstate of the system. This can be done by establishing a protocol of allowed single (sorting some spin to flip or a new position for some molecule) or collective (like Wolff algorithm \cite{PhysRevLett.62.361}) movements. In this work we used single-spin flip. Within the defined protocol, if the system is in a microstate with energy $E$, the movements can be labelled as: 
\begin{eqnarray}
\mbox{type I : } & E \longrightarrow E - \Delta E\nonumber \\
\mbox{type II : } & E \longrightarrow E + \Delta E\nonumber \\
\mbox{type III : } & E \longrightarrow E \nonumber,
\end{eqnarray}
where $\Delta E > 0$ and does not always corresponds to the same value. That is, movements which decrease, increase or does not change, respectively, the energy of the system. One naive attempt would be to accept sorted movements with equal probability. It would make the energy RW to be biased, because $g(E)$ is a monotonically growing function. In the positive temperature region there are more type II movements than type I, which would make the RW to sample configurations with higher energies until $\langle E\rangle_{T\rightarrow\infty}$.

The WLS \cite{PhysRevLett.86.2050, PhysRevE.64.056101} consists of performing a set of energy RW, in such a way that the acceptance of the new configuration sorted is inversely proportional to the density of states associated with its energy. This way, the probability of type II movements are smaller, preventing the RW dynamics to go to higher energies. During each RW, the algorithm saves in a histogram $H(E)$ the number of sampled configurations with energy $E$. The density of states of each new sampled energy is changed by a factor $f_i$, $g(E)\rightarrow g(E)\times f_i$. This factor is mildly reduced during the simulation, so that at the end of the algorithm execution it is very close to one. This ensures a uniform sampling (or a flat histogram) of all energy levels, from which $g(E)$ can be determined. 

In the beginning of the simulation $g(E)$ is not known, so the best initial guess is to take $g(E)=1$, $\forall E$. From a configuration with energy $E_1$, a new random configuration with energy $E_2$ is reached with probability
\begin{equation}
p(E_1\rightarrow E_2) = \min\left(\frac{g(E_1)}{g(E_2)},1\right).
\label{wl_weight}
\end{equation}
This equation implies that whenever $g(E_2)\leq g(E_1)$ the new configuration is accepted. Otherwise, we sort a uniformly distributed random number $r$, between 0 and 1, and accept the new state if $r\leq g(E_1)/g(E_2)$. If the new state is accepted, we update the histogram, $H(E_2) = H(E_2)+1$, and the density of states, $g(E_2) = g(E_2)\times f_i$. If it is not, 
$H(E_1) = H(E_1) +1$ and $g(E_1) = g(E_1)\times f_i$. This evaluation of the density of states tends to produce numbers too large to be represented by double precision numbers in the computer. Thus, we work with the natural logarithm of the density of states $\ln [g(E)]$. This way the density of states is updated as $\ln[g(E)] = \ln[g(E)] + \ln f$ and the probability of transition is given by 
\begin{equation}
p(E_1\rightarrow E_2) = \min\left(\exp\left[\ln g(E_1) - \ln g(E_2)\right],1\right).
\label{wl_weight2}
\end{equation}

We choose the initial value of the modification factor to be $f_0=2.71828...$ $=e$, and it is updated whenever $H(E)$ is considered flat. In practice we check the histogram for flatness every 10 000 MC sweeps (each sweep corresponds to $N$ trial movements), and it is considered flat when $H(E) > x\%\times\langle H\rangle$ for all energy levels. Where $\langle H\rangle$ is the mean value of $H(E)$ and $x\%$ is some percentage of how plan $H(E)$ is (in general $80\leq x\leq 90$). If this condition is fulfilled, $\ln[g(E)]$ will have converged to the exact value with precision proportional to $\ln f$ and we proceed to the next stage by updating $f_{i+1}=\sqrt{f_i}$ and reinitializing the visitation histogram, $H(E) =0\,\,\forall E$. The conventional execution of the WLS algorithm is halted when 
$\ln f \sim 10^{-8}$, which happens when $f_i = f_{26}$. At the end of the simulation one can use Eq.~(\ref{mediacan2}) to evaluate $\langle Q\rangle_T$, with the microcanonical averages $\langle Q(E)\rangle$. The microcanonical averages are calculated with the flat histogram $h(E)$, accumulated during the whole execution of the algorithm.

\begin{figure}[!htb]
\begin{center}
\includegraphics[width=0.5\textwidth]{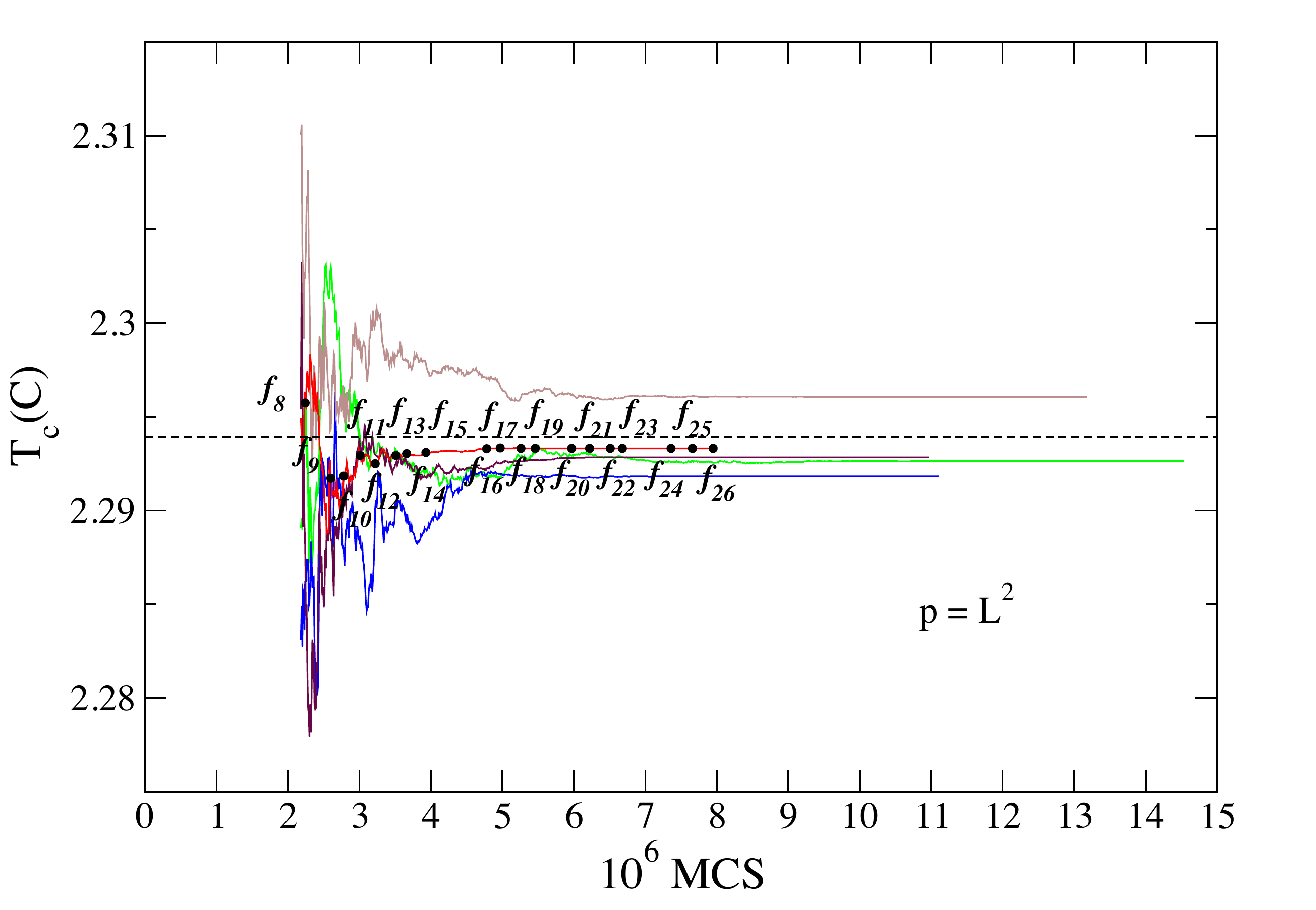}
\caption{Evolution of the temperature associated with the peak of $C_v(T)$ for five independent samples, during the WLS, begining from $f_8$ and using $\%x = 85$. The dots indicates the update of $f$ and the horizontal dashed line corresponds to $T^{the}_c(L=32) = 2.29392979$ obtained from data in Ref.~\cite{IsingExactMathematica}.}
\label{TcCvEvol_L32}
\end{center}
\end{figure}

Defined this way the WLS method arbitrates a stopping point for the evaluation of $g(E)$ and introduces correlations by sampling from successive trial configurations. In Ref.~\cite{PhysRevE.72.025701} the authors prove the convergence and discuss the systematic error caused by adjacent records in $h(E)$, which can be controlled by reducing $f$, and also by increasing the number of trial configurations $p$, between records. A practical computational study of the convergence and accuracy was recently presented in \cite{PhysRevE.85.046702}. By investigating the behaviour of the peak of response functions such as the specific heat,
\begin{equation}
C = \frac{\left\langle E^2\right\rangle_T - \left\langle E\right\rangle^2_T}{T^2},
\label{Cv}
\end{equation}
and magnetic susceptibility,
\begin{equation}
\chi = L^2\frac{\left\langle m^2\right\rangle_T - \left\langle\vert m\vert\right\rangle^2_T}{T},
\label{Xt}
\end{equation} 
where $E$ and $m$ are the energy and magnetization, respectively, calculated during the simulation using Eq.~(\ref{mediacan2}), one can follow the convergence of the algorithm. Figure \ref{TcCvEvol_L32} shows the evolution of temperature related to the peak of $C$ calculated for the 2D Ising ferromagnet in a square lattice with $L=32$, for five independent runs as a function Monte Carlo sweeps. We verify that there is no reason to continue the simulation run beyond $\ln f_{23} = 1,1921\times 10^{-7}$, because the density of states converged. The convergence occurs for different values of $T_c(C)$, around $T^{the}_c(L=32) $, for each sample. To establish a criteria to halt the simulation, the author in Ref.~\cite{PhysRevE.89.043301} suggests to stablish a threshold for the difference between $T_c(0)$, associated with the peak of the specific heat when $H(E)$ is considered flat, and $T_c(t)$ calculated every time $H(E)$ is tested for flatness at the current modification factor $f_i$. When this difference is bellow such predefined threshold, the simulation is halted.

\begin{figure}[!htb]
\begin{center}
 \includegraphics[width=0.5\textwidth]{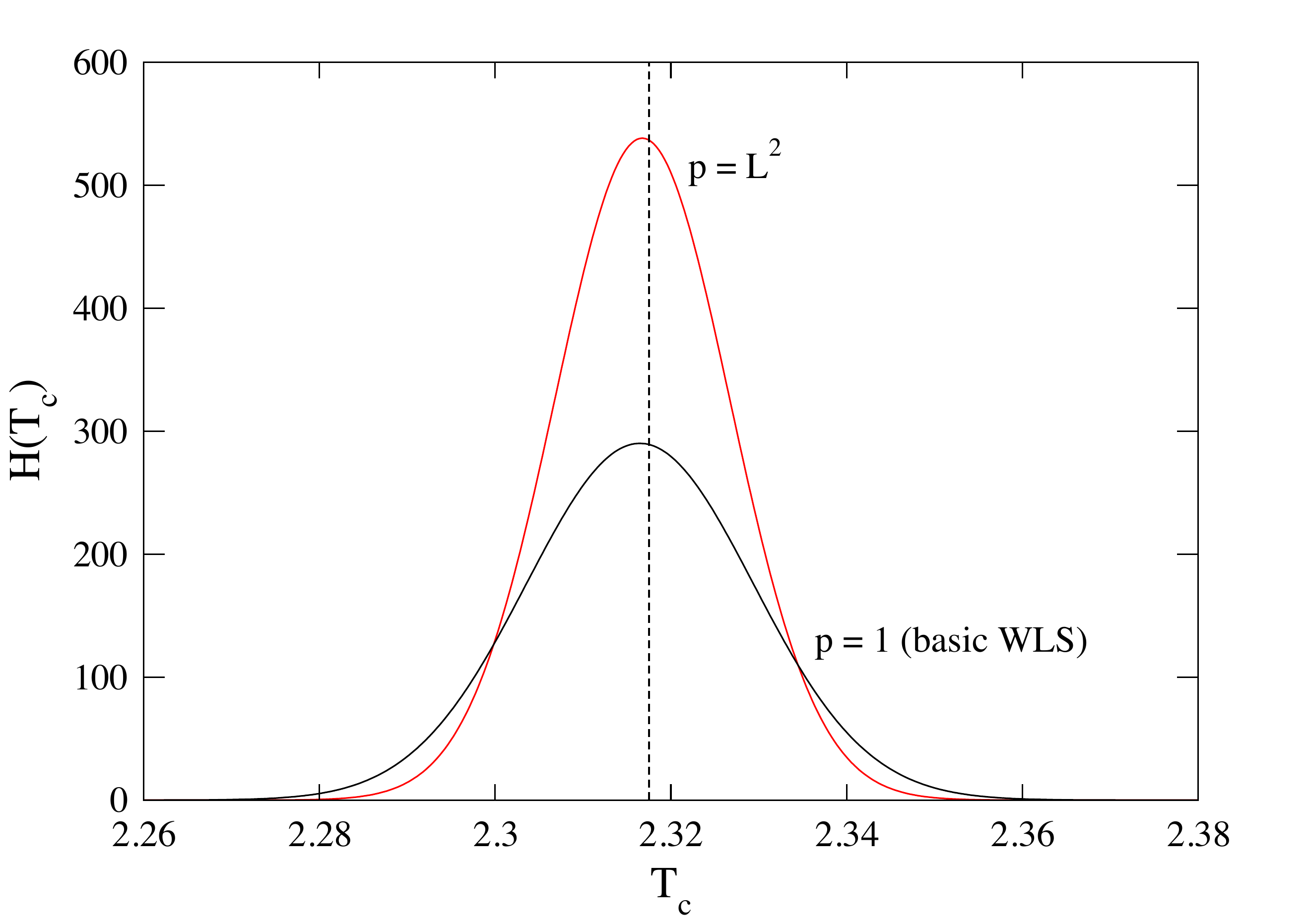}
 \caption{Best-fit Gaussians for the histogram of $T_c(C)$ of the simulation for 50 000 independent runs for $f$ up to $\ln f = 10^{-4}$ and $x=85\%$ . The vertical line corresponds to $T^{the}_c(L=16) = 2.3175165$ obtained from Ref.~\cite{PhysRevLett.76.78}.}
 \label{HisTcL16}
 \end{center}
\end{figure}

To improve the accuracy of the WLS we study the effect of taking $p$ trial sweeps between records in $h(E)$. As Fig.~\ref{HisTcL16} shows, the dispersion of the temperature associated with the peak of the specific heat is reduced by increasing $p$. Then, we adopt in our simulations $p=L^2$, i.e., the density of states is updated every $L^2$ trial configurations, which is the standard MC step. In Ref.~\cite{PhysRevE.85.046702}, the authors also demonstrates that the microcanonical averages of $\langle m(E)\rangle$, for each energy channel, vary in the initial (large $f$) phase of the WLS and follow steady values for $f=f_{micro}$. Therefore, to improve the accuracy of the microcanonical averages, one should only take samples for $\langle Q(E)\rangle$ after $f_{micro}$.

Another method to calculate $g(E)$ used in this work is the Broad Histogram Method \cite{BHM_BJP,BHM_relation}. This method is not related with any specific protocol to sample microstates, but with the microcanonical averages of two quantities, $N_{up}$ and $N_{dn}$. These quantities measure the total number of possible movements that can be made (but not necessarily performed) from a state with energy $E$ that increases and decreases, respectively, the system's energy by a fixed amount $\Delta E$ (in Ref.~\cite{BHMC}, the method is extended to systems with continuous energy spectra). The sampling protocol is defined mathematically through the transition matrix $\mathfrak{M}(S,S')$, whose elements are 1 (allowed movement) and 0 (forbiden movement). The only criteria to be fulfilled by the dynamics, to use the BHM, is microreversibility. Such criteria is satisfied if $\mathfrak{M}(S,S')$ is symmetric: if the transition $S \rightarrow S'$ is permitted, the transition $S' \rightarrow S$ also is, regardless the probability of each one to occur.

Given a microstate $s_i$ with energy $E$, there are $N_{up}^{+\Delta E}\left[E(s_i)\right]$ movements that, if performed, increase the energy by $\Delta E$. Just as there are also $N_{dn}^{-\Delta E}\left[E'(s_j)\right]$ possible movements from a microstate $s_j$, with energy $E'=E+\Delta E$, that decreases the energy by the same amount $\Delta E$. Due to the quoted microreversibility criteria, we can write:
\begin{equation}
\sum_{\lbrace s_i(E)\rbrace} N_{up}^{\Delta E}(E) = \sum_{\lbrace s_j(E+\Delta E)\rbrace} N_{dn}^{-\Delta E}(E+\Delta E),
\label{NupNdn}
\end{equation} 
where the sums are performed for all microstates $s_i$ ($s_j$) with energy $E$ ($E+\Delta E$). Using Eq.~(\ref{mediamicro}) we can rewrite this equation as,
\begin{equation}
\langle N_{up}(E)\rangle g(E) = \langle N_{dn}(E+\Delta E)\rangle g(E+\Delta E).
\label{BHM_relation}
\end{equation}
This is the fundamental relation of the Broad Histogram method. It allows us to evaluate the density of states, recursively, from the \textbf{macroscopic} $N_{up}$ and $N_{dn}$ averaged during the execution of the sampling dynamics (whatever is it, maybe based in the same protocol maybe another, for instance WLS as adopted here), for all energy levels. To use Eq.~(\ref{BHM_relation}), one have to adopt an initial guess for the initial energy, $g(E_0)$ and $N_{up}(E_0)$ [$g(0)=2$ and $N_{up}^{+4}(0)=N$ are the exact values for the ground state of the Ising ferromagnet, e.g.] and then use: 
\begin{equation}
g(E_0+\Delta E) = \frac{\langle N_{up}(E_0)\rangle}{\langle N_{dn}(E_0+\Delta E)\rangle}g(E_0),
\end{equation}
to scan the calculation through all energy axis.

The approach used in this article implements the WLS to get an estimation for the density of states and the microcanonical averages $\left\langle Q(E)\right\rangle$. Among the usual histograms of the magnetization $m(E)$, and its even powers $m^2(E)$ and $m^4(E)$, we also calculate the histograms of $N_{up}^{+4}(E)$, $N_{up}^{+2}(E)$, $N_{dn}^{-2}(E)$ and $N_{dn}^{-4}(E)$. To improve accuracy, we use a set of independent runs (one hundred for each lattice size) and the criteria established in Refs.\cite{PhysRevE.85.046702, PhysRevE.89.043301}. 

\section{Results and discussion}
In order to compare both approaches to calculate the density of states, we use the $L\times L$ Ising model with nearest neighbour interaction, 
\begin{equation}
\mathcal{H} = -J\sum_{<ij>}\sigma_i\sigma_j,
\label{IsingHamiltonian}
\end{equation}
which stands as a benchmark for new theories and simulation methods. It is auspicious to use this model because it has an exact solution for infinite systems \cite{PhysRev.65.117}, allowing finite size scaling analysis results comparison, as well as for finite systems \cite{PhysRevLett.76.78}, where the author uses a MATHEMATICA program \cite{IsingExactMathematica} to expand the partition function and provide exact solution. We simulated lattice sizes up to $L=128$ with periodic boundary conditions, which is the maximum feasible size we can evaluate with the mentioned MATHEMATICA program. In our simulations the energy, $E=0,4,6,8, ...,2L^2$, is accounted as the number of unsatisfied bonds ($E =0$ addresses the state where all spins are aligned and $E= 2L^2$ the N\' eel's state energy) and expressed in terms of the energy density $e = E/2L^2$. So, in this way, the energy is in units of $J$ and the energy axis is reflected in relation to Eq.~(\ref{IsingHamiltonian}). We limited our attention to the region where the density of states is an increasing function of the energy, i.e. the positive-temperature region. 

\begin{figure}[!htb]
\begin{center}
 \includegraphics[width=0.5\textwidth]{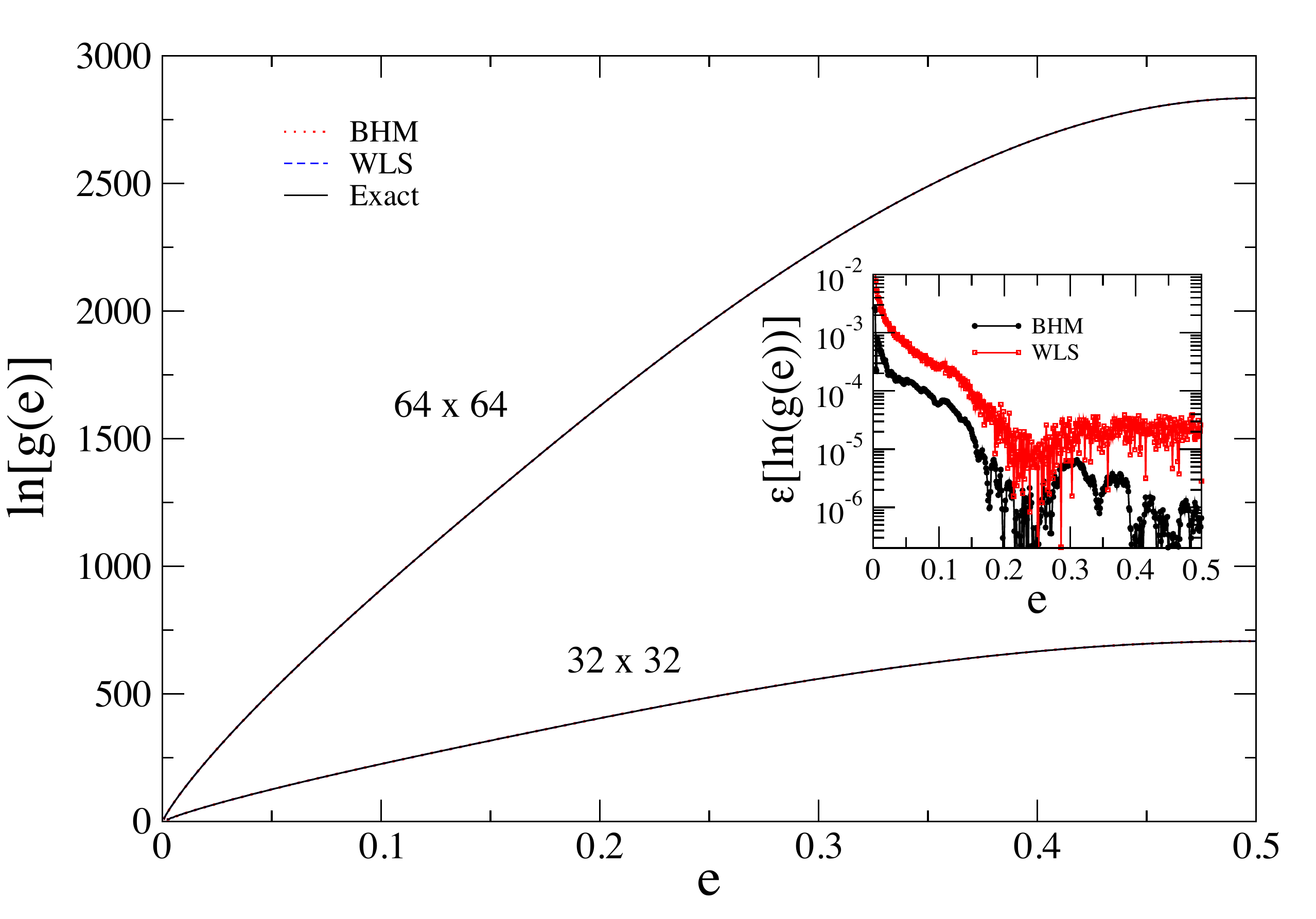}
 \caption{Comparison of the density of states obtained by the WLS, BHM and the exact result calculated with Ref.~\cite{PhysRevLett.76.78}. Due to indistinguishability between the exact and the simulation results, the inset shows the relative errors $\varepsilon \left[\ln(g(e))\right]$ for the $32 \times 32$ case.}
 \label{Dif_DOS_32_64}
 \end{center}
\end{figure}

Figure~\ref{Dif_DOS_32_64} shows the densities of states for two lattice sizes, $L=32$ and $64$, obtained by the random walk performed in the WLS, the BHM and the exact calculation provided by \cite{IsingExactMathematica}. It is not possible to distinguish the simulation results from the exact ones. Thus, to magnify the difference between both methods, we also show the relative error $\varepsilon(X)\equiv \vert X_{sim} - X_{the}\vert/X_{the}$. The logarithm of the density of states is not the best way to verify the precision difference between the methods. As the inset in Fig.~\ref{Dif_DOS_32_64} shows, the mean relative error is dominated by small energies, where the error is larger. As normally one is interested in thermal averages such as the Eq.~\ref{mediacan2}, to verify the accuracy of the methods table~\ref{MeanErr} exhibits the relative error of the quantity $\sum_{E} g(E)\exp\left(-E/T_c\right)$, where $T_c$ is the critical temperature. 

\begin{table}[H]
\centering
\caption{Relative errors of $\sum_{E} g(E)\exp\left(-E/T_c\right)$ for WLS and BHM for all lattice sizes simulated and the total Monte Carlo steps per independent sample for each size.}
\label{MeanErr}
\begin{tabular}{cccc}
\hline
L   & WLS      & BHM      & MCS                            \\ \hline
24  & 4.304$\%$ & 1.462$\%$ & $3.5\times 10^5$ \\
32  & 3.739$\%$ & 0.806$\%$ & $8.5\times 10^5$\\
48  & 2.150$\%$ & 0.090$\%$ & $3.0\times 10^6$ \\
64  & 1.580$\%$ & 1.015$\%$ & $6.9\times 10^6$ \\
96  & 7.099$\%$ & 0.604$\%$ & $2.9\times 10^7$ \\
128 & 2.864$\%$ & 0.994$\%$ & $8.4\times 10^7$ \\ \hline
\end{tabular}
\end{table}

\begin{figure}[!htb]
\begin{center}
 \includegraphics[width=0.5\textwidth]{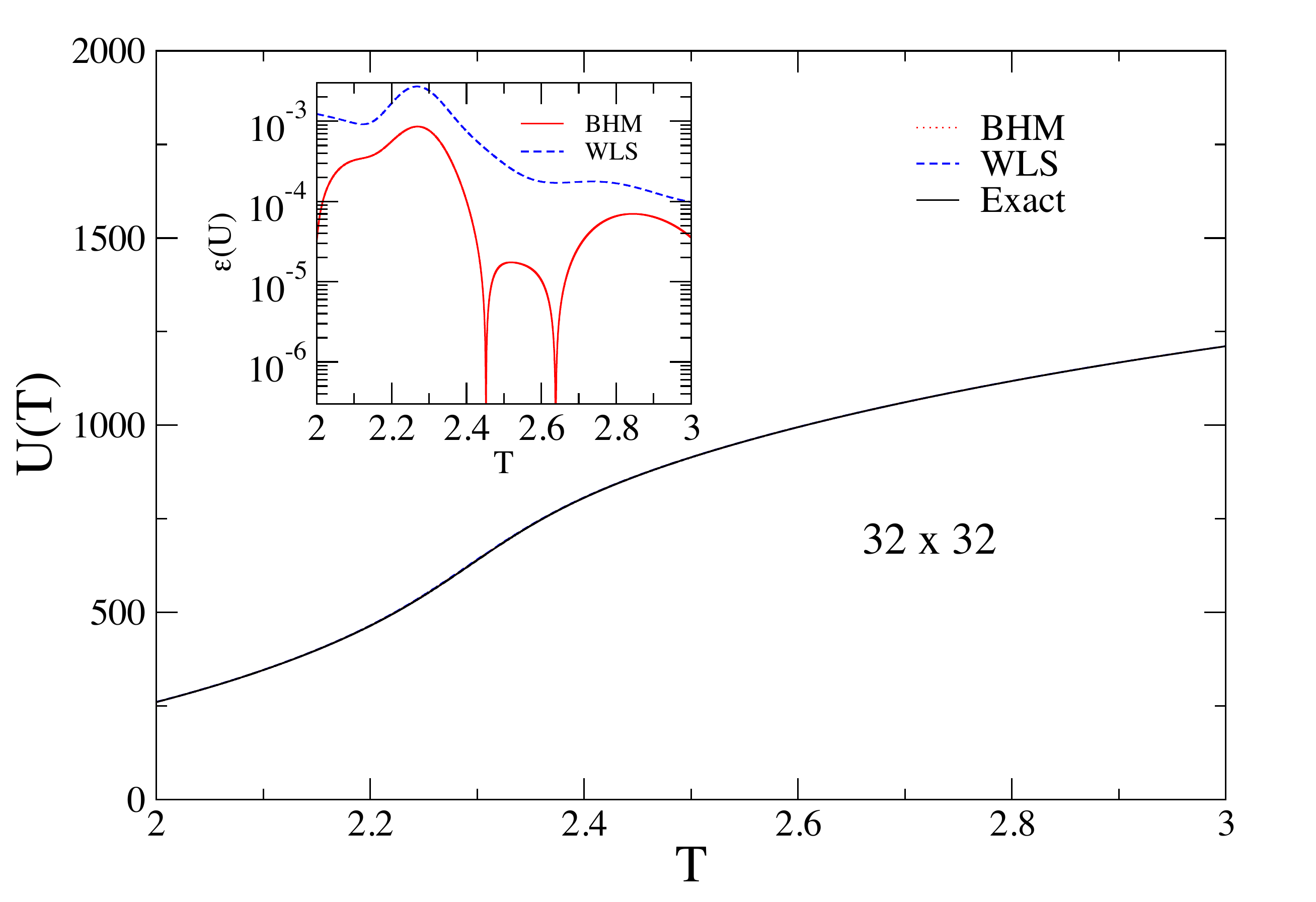}
 \caption{Comparison of the thermal energy obtained by the BHM, WLS and the exact result calculated with Beale's exact result \cite{PhysRevLett.76.78}. Once again, due to indistinguishability between the exact and simulation results, the inset shows the relative errors $\varepsilon(U)$ for the $32 \times 32$ case.}
 \label{Dif_E_32_64}
 \end{center}
\end{figure}

With the densities of states obtained we can use Eq.(\ref{mediacan2}) to describe the thermodynamic properties of the system. It is worthwhile to emphasize that a single simulation provides both estimations for the density of states, the microcanonical averages are the ones obtained by the WLS and no other dynamics was involved in our simulations. In Fig.~\ref{Dif_E_32_64}, we compare the thermal energy as a function of temperature, $U(T) = \left\langle E\right\rangle_T =\sum_{\mathbf{E}} g(E)E\exp(-E/T)/\sum_{\mathbf{E}}g(E)\exp(-E/T)$, provided by both techniques for a temperature interval $T\,\in\,[2,\,3]$. The agreement of both methods with the exact result is very good and, once more, the BHM is more precise and displays a relative error at least one order of magnitude smaller than the WLS for the $32\times 32$ case. The advantage of BHM is larger yet for larger lattices, since $N_{up}$ and $N_{dn}$ are macroscopic quantities.

\begin{figure}[!htb]
\begin{center}
 \includegraphics[width=0.5\textwidth]{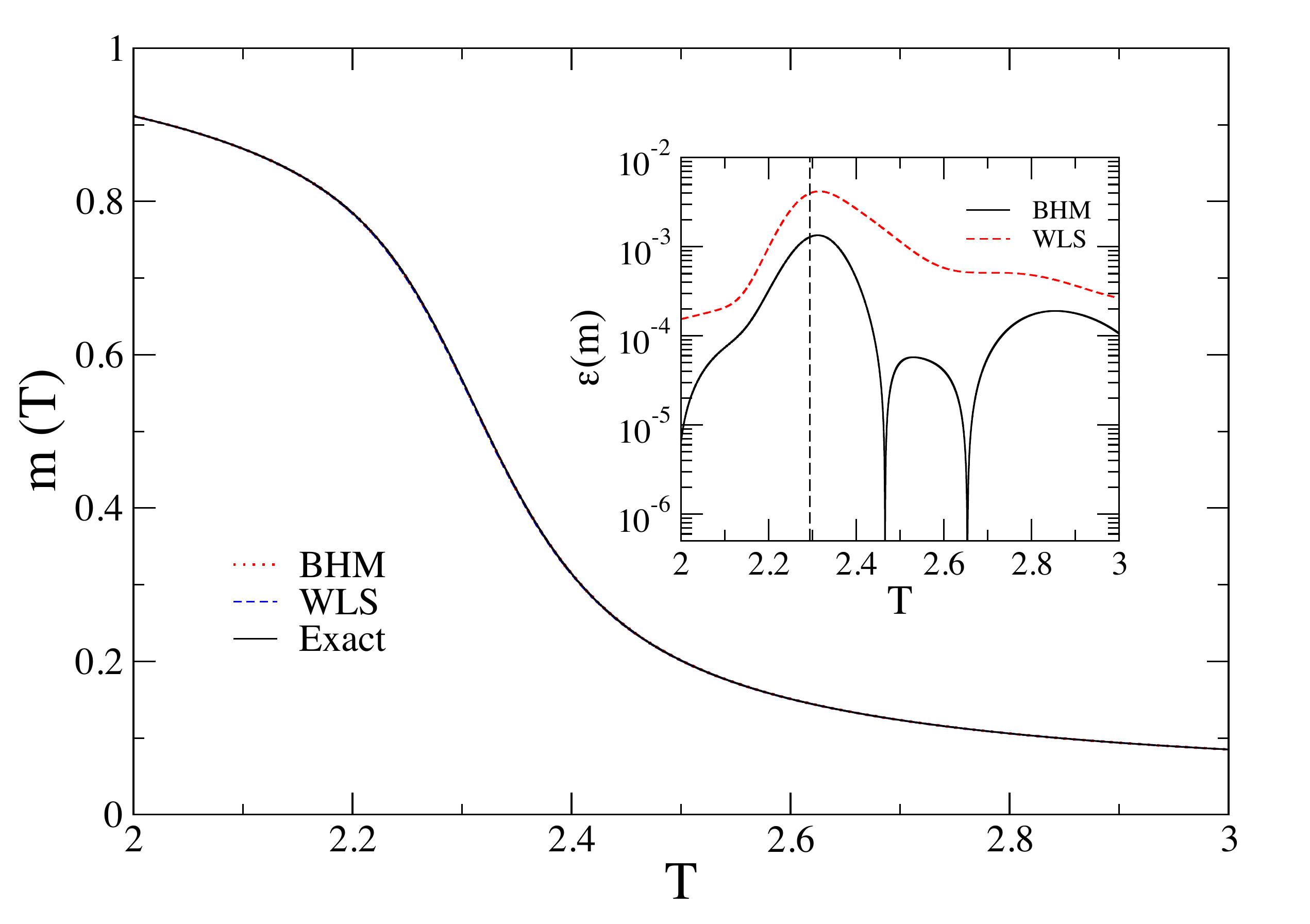}
 \caption{Comparison of the magnetization as function of the temperature obtained by the BHM, WLS and the Beale's exact result (using the microcanonical averages from the WLS walk). The difference between the exact and simulation results is not visible, so the inset shows the relative errors $\varepsilon(m)$ for the $L=32$ case. The vertical dashed line in the inset corresponds to $T_c(L=32) = 2.29392979$, obtained using the exact data from Ref.~\cite{IsingExactMathematica}.}
 \label{Dif_M_32_64}
 \end{center}
\end{figure}

To compare the magnetization as a function of the temperature as shown in Fig.~\ref{Dif_M_32_64}, we used the microcanonical averages $\left\langle m\right\rangle_E$ obtained from the WLS samples for the $L=32$ case. The difference between the exact and simulation results is not visible, so the inset shows the relative errors $\varepsilon(m)$. Even with the statistical error introduced by the microcanonical averaging of the WLS random walk, the BHM is again one order of magnitude more precise in the critical region. 

\begin{figure}[!htb]
\begin{center}
 \includegraphics[width=0.5\textwidth]{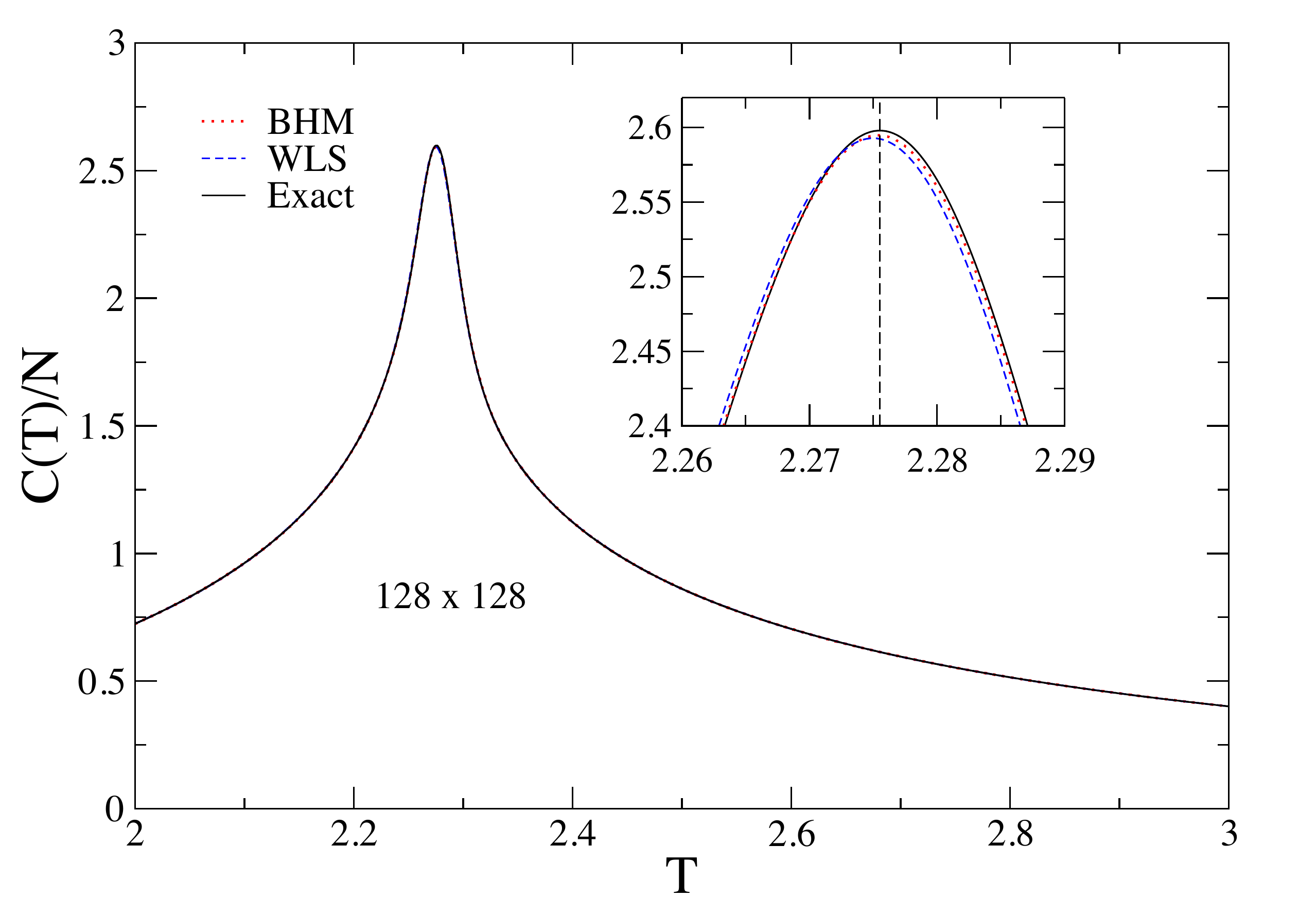}
 \caption{Specific heat for the $128\times 128$ Ising ferromagnet. We compare both methods with the exact result in the temperature range $T\in [2,3]$ and the inset magnifies the peak of the specific heat, where commonly the agreement is frail for most of the simulation methods. The vertical dashed line in the inset corresponds to the critical temperature obtained from Ref.~\cite{PhysRev.185.832}, $T_c^{the}(L=128)=2.2755091$.}
 \label{Cv_T}
 \end{center}
\end{figure}

A more rigorous test of accuracy of the density of states is the calculation of response functions such as the specific heat and the magnetic susceptibility. The specific heat provides a pinpoint check for the precision, since it has an exact solution obtained by Ferdinand and Fisher in \cite{PhysRev.185.832}, and it is not related with the microcanonical averages involved in the calculation of the susceptibility. Our simulation results for both methods are compared with the exact result in Fig.~\ref{Cv_T} for $L=128$. The agreement of both methods with the exact result is excellent, but the BHM provides a more accurate description of the density of states. The inset highlights the critical region, where the specific heat has a blow up located at $T_c^{the}(L=128)=2.2755091$ with height $C_{max}(L=128)/N = 2.538417$. For the BHM the blow up occurs at $T_c^{BHM}(L=128)=2.27538(31)$ and for the WLS at $T_c^{BHM}(L=128)=2.27498(08)$, where the parentheses indicates the error bars corresponding to the last two digits. 

\begin{figure}[!htb]
\begin{center}
 \includegraphics[width=0.5\textwidth]{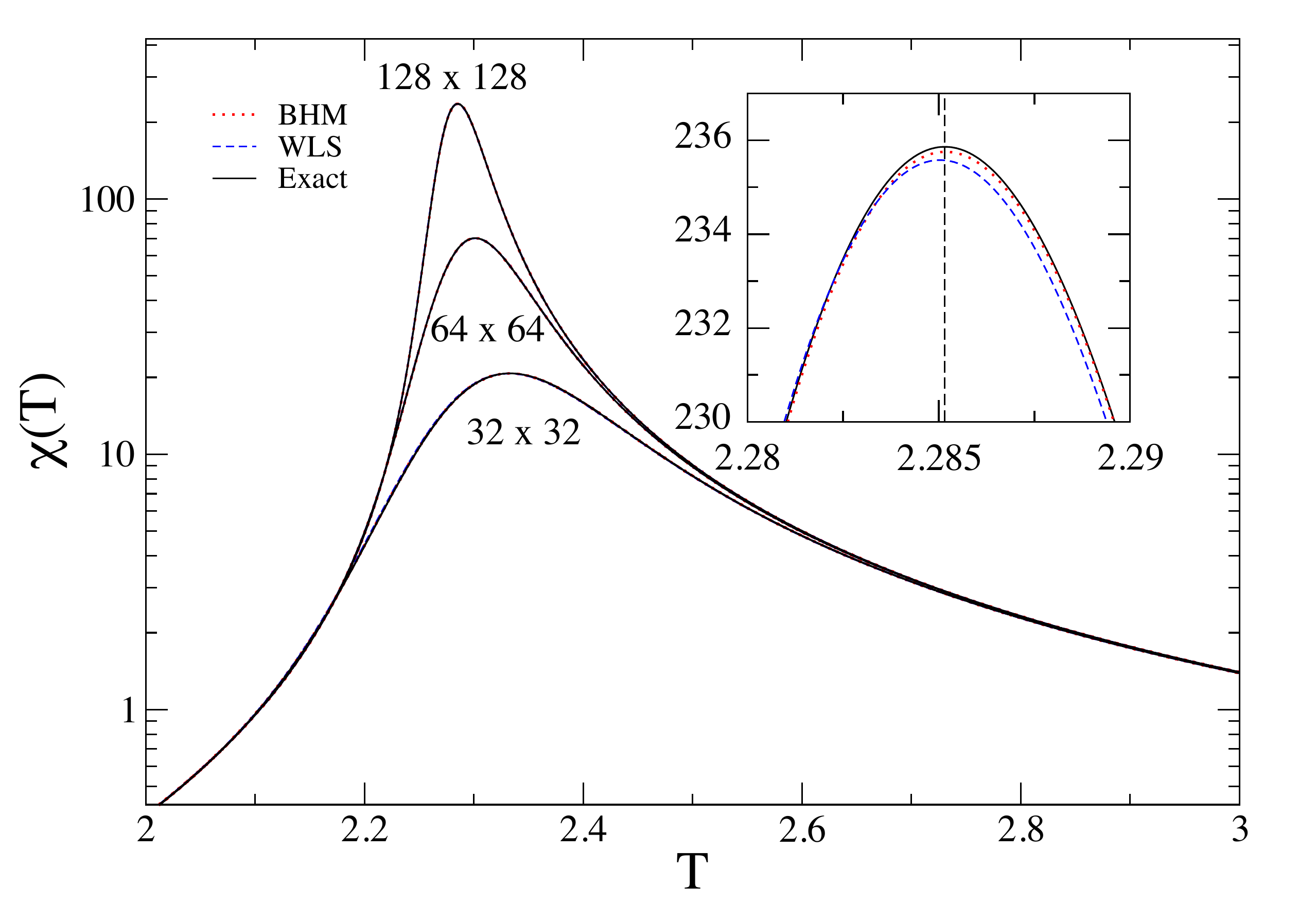}
 \caption{Mono-log scale magnetic susceptibility for the $128\times 128$, $64\times 64$ and $32\times 32$ Ising ferromagnet. The vertical dashed line in the inset corresponds to the temperature associated with the peak of susceptibility, $T_c^{the}(L=128)=2.2851601$, obtained from~\cite{PhysRevLett.76.78} and using the microcanonical averages $\left\langle m\right\rangle_E$ and $\left\langle m^2\right\rangle_E$, accumulated during the set of WLS performed for each lattice size.}
 \label{X_T}
 \end{center}
\end{figure}

The susceptibility also stands as test of the precision of both methods, although it depends on the precision of the microcanonical averages evaluation during the WLS. Figure \ref{X_T} compares both methods with the exact result from Ref.~\cite{PhysRevLett.76.78}, in the temperature range $T\in [2,3]$. In the inset we magnify the peak of the susceptibility for the $128\times 128$ case, and one can see that the agreement with the exact case is better for the BHM. The random walk performed by the WLS is only a Markovian process if $p$ [the number of trial movements between records in $H(E)$] is taken arbitrarily large, as proved in Ref.~\cite{PhysRevE.72.025701} and corroborated by the results in Ref.~\cite{PhysRevE.85.046702}. During the sampling process the density of states oscillates when $f_i$ is large (in the initial phase of the simulation) and, as the $f_i$ becomes small (by the end of the simulation), it converges to a value in the vicinity, apart $\sqrt{\ln f}$, of the exact one as Ref.~\cite{PhysRevE.89.043301} shows. 

Even when all strategies to improve the convergence are applied, the BHM is more accurate. The main reason why the Broad Histogram Method is more accurate, for all thermodynamic quantities and lattice sizes, than the Wang-Landau Sampling is related with the exact nature~\cite{BHM_relation} of Eq.~(\ref{BHM_relation}). Furthermore, the BHM evaluates the density of states by means of the \textbf{macroscopic} quantities, $N_{up,dn}^{\pm\Delta E}$: each new averaging state contributes with a macroscopic value to the averaging histograms, instead of just incrementing by unit the number of visits to the corresponding energy channel. All measures in order to improve the WLS, will also improve the microcanonical averages and, by this mean, will improve the accuracy of the BHM estimation of the density of states.

As a final test of accuracy we can apply finite-size scaling analysis~\cite{PhysRevLett.28.1516} in order to describe the critical behaviour of the Ising ferromagnet, which has the critical exponents exactly known. Through the definition of the free energy, one can obtain the zero-field scaling relations for the magnetization,
\begin{equation}
m\approx L^{-\beta/\nu}\mathcal{M}(tL^{1/\nu}),
\label{ScaleMag}
\end{equation}
and for the magnetic susceptibility,
\begin{equation}
\chi\approx L^{\gamma/\nu}\mathcal{X}(tL^{1/\nu}).
\label{ScaleSus}
\end{equation}

\begin{figure}[!htb]
\begin{center}
 \includegraphics[width=0.5\textwidth]{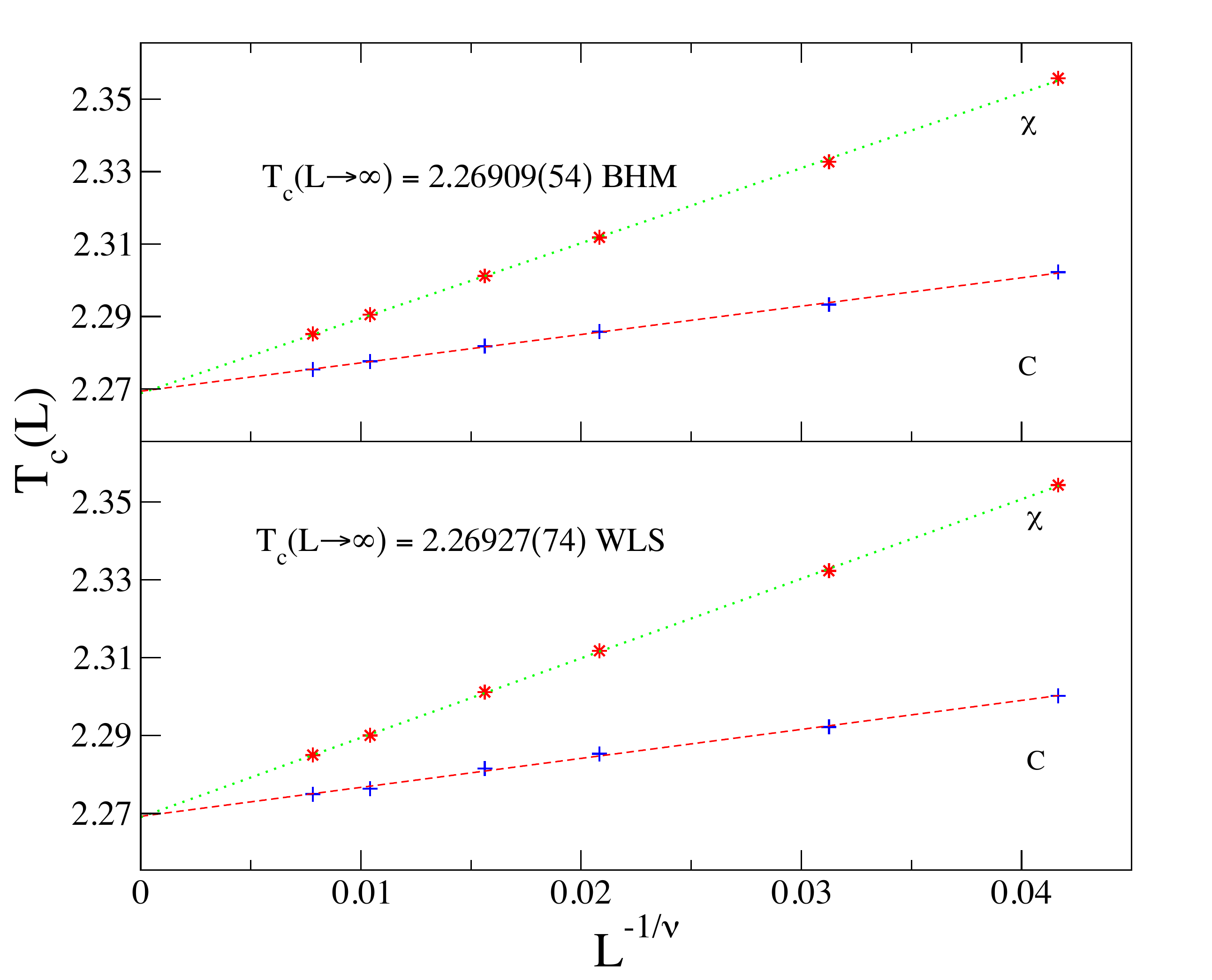}
 \caption{Size-dependent critical temperature associated with the peak of the susceptibility and specific heat. The upper panel displays the results obtained with the BHM and the lower panel the WLS one, assuming $\nu=1$ for both cases.}
 \label{Tcinf_CvXt_BHM_WL}
 \end{center}
\end{figure}

The size-dependent critical temperature, which is associated with the peak of the specific heat and susceptibility, scales asymptotically as 
\begin{equation}
T_c(L) \approx T_{c}^{L\rightarrow\infty} + q\,L^{-1/\nu},
\label{ScaleTemp}
\end{equation}
where $q$ is a quantity-dependent constant. This relation allows us to get an estimate of the critical temperature given by both methods, as shown in Fig.~\ref{Tcinf_CvXt_BHM_WL} where we assumed $\nu=1$. We can see that the linear fits for $C$ and for $\chi$ converges to $T_c^{\infty}$ as $L^{-1/\nu}=0$ and that the $T_c^{\infty}$ associated with both quantities are slightly different. The estimate indicated in Fig.~\ref{Tcinf_CvXt_BHM_WL} corresponds to the mean value of both fits.

\begin{figure}[!htb]
\begin{center}
 \includegraphics[width=0.5\textwidth]{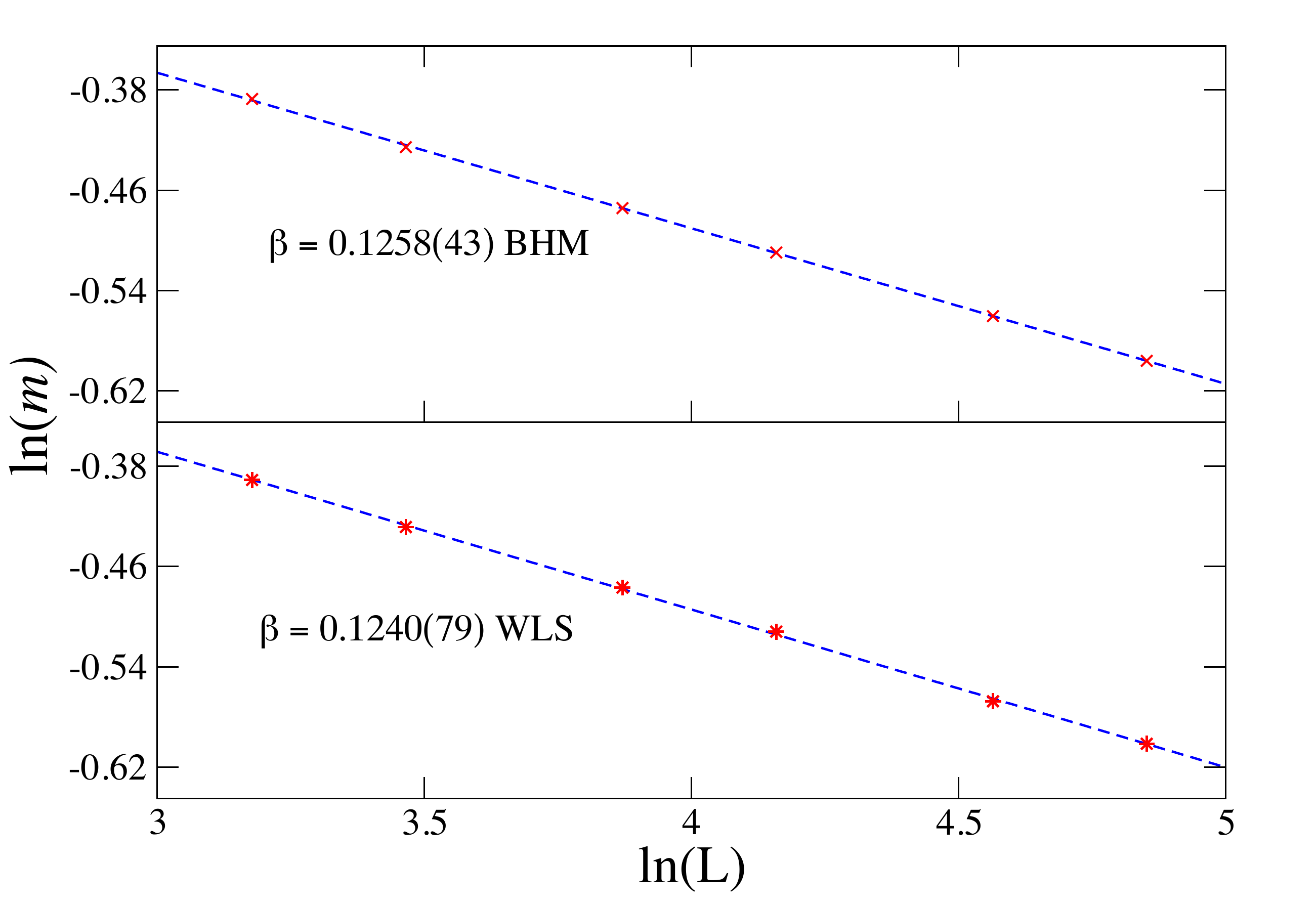}
 \caption{Log-log plot of the finite lattice-size dependence of the magnetization using $T_c=2.26909$ for the BHM (top) and $T_c=2.26927$ for the WLS (bottom).}
 \label{beta_BHM_WL}
 \end{center}
\end{figure}

\begin{figure}[!htb]
\begin{center}
 \includegraphics[width=0.5\textwidth]{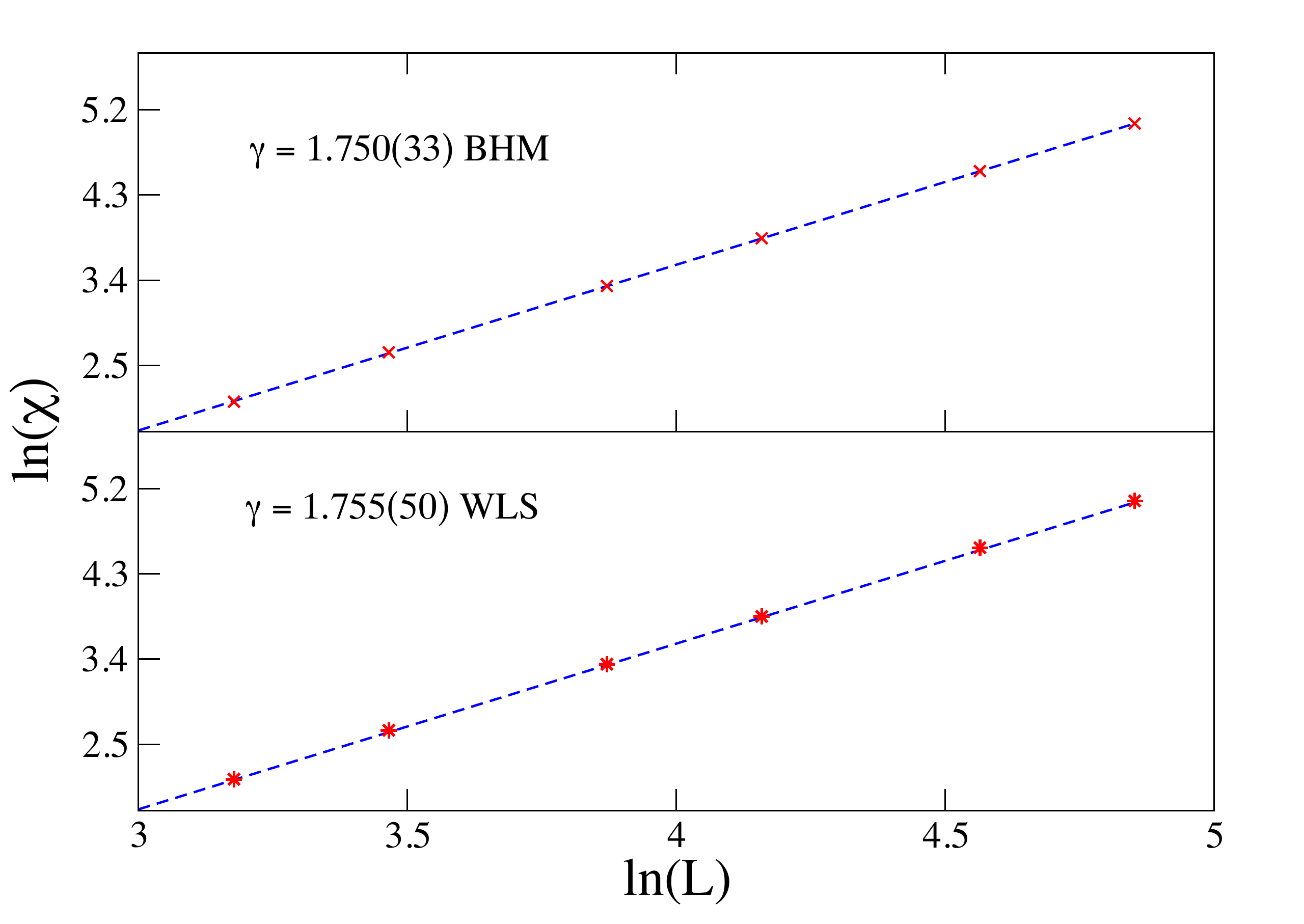}
 \caption{Log-log plot of the finite lattice-size dependence of the susceptibility using $T_c=2.26909$ for the BHM (top) and $T_c=2.26927$ for the WLS (bottom).}
 \label{gamma_BHM_WL}
 \end{center}
\end{figure}

Now that we have an estimate of $T_c$ for both methods, we can calculate the magnetization and susceptibility at $T_c^{BHM}$ and $T_c^{WLS}$ in order to estimate $\beta$ and $\gamma$, respectively. Equation (\ref{ScaleMag}) states that the finite-size magnetization scales asymptotically with $L^{-\beta/\nu}$, when the temperature is nearby $T_c$. So we use the slope of the log-log scale graph shown in Fig.~\ref{beta_BHM_WL} to obtain $\beta$ for both methods. We adopted the very same process for the susceptibility to obtain $\gamma$, using Eq.~(\ref{ScaleSus}). In Fig.~\ref{gamma_BHM_WL}, we show the lattice-size dependence of the peak of susceptibility in \textit{log-log} scale and we obtain, through a linear fit, the critical exponent $\gamma$. 

\begin{figure}[!htb]
\begin{center}
 \includegraphics[width=0.5\textwidth]{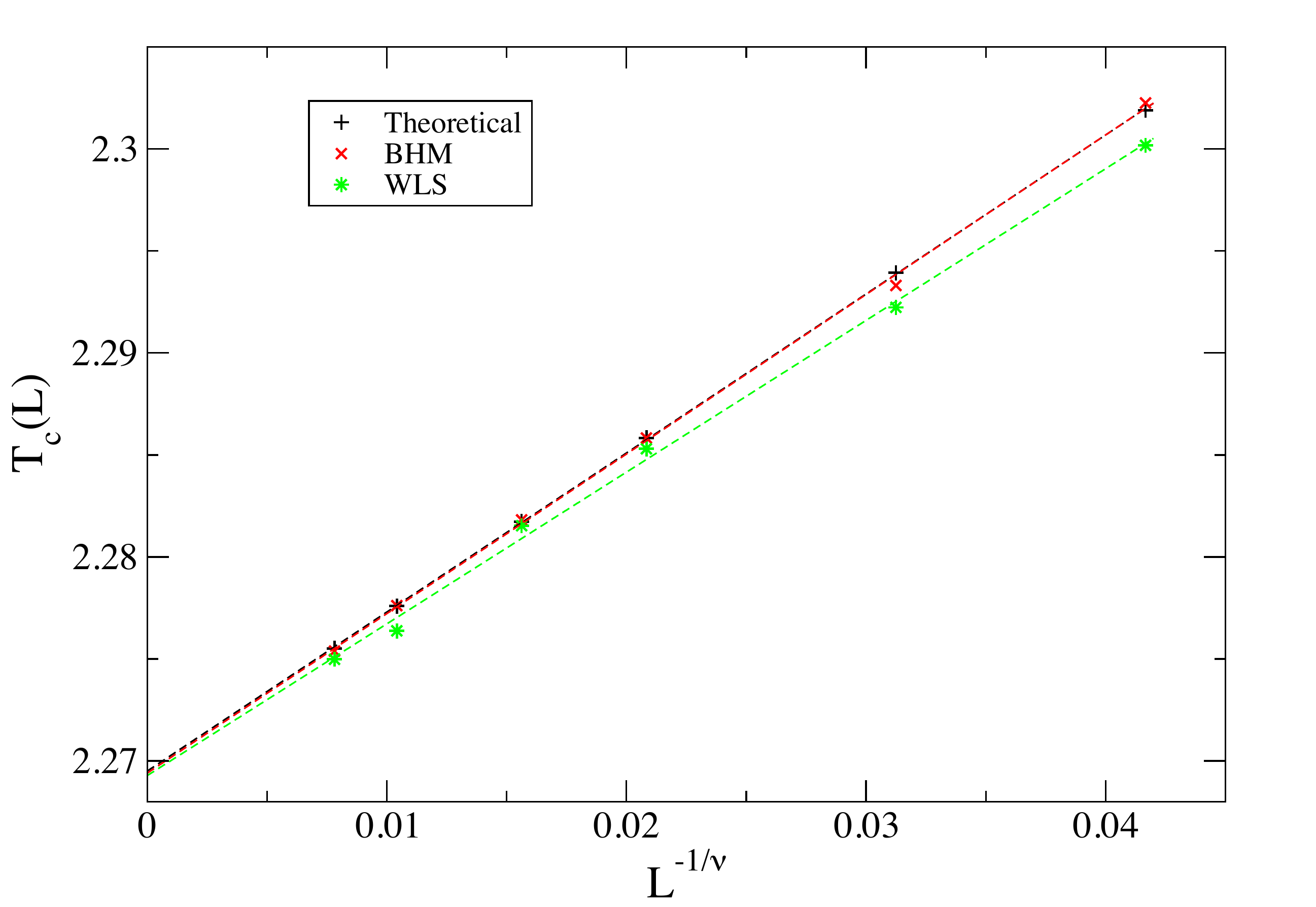}
 \caption{Size-dependent critical temperature associated with the peak of the specific heat. Direct comparison between the exact result with BHM and WLS, assuming $\nu=1$ for both cases.}
 \label{TcCv_Beale_BHM_WL}
 \end{center}
\end{figure}

The thermodynamic limit extrapolations from finite-size data, as the ones we used to compare with the exact results, would require higher order terms in Eq.~(\ref{ScaleTemp}), which are unknown in finite-size scaling theory. A direct solution is to perform simulations with larger system sizes that would make such higher order terms negligible. However this is not our aim here, which is to compare the BHM and the WLS with the exact results available for finite system sizes, as Fig.~\ref{TcCv_Beale_BHM_WL} shows.

\section{Conclusions}

By extensive comparison with exact results for the thermodynamic properties of the two-dimensional Ising model, we have demonstrated that when the density of states is calculated with macroscopic quantities, as done by the Broad Histogram Method, there is a gain in accuracy within the Wang-Landau sampling. Even when the convergence of the WLS is carefully controlled, by following the temperature associated with the peak of the specific heat, and when we take a set of independent runs, the BHM is a more precise way of calculating the density of states. 

It is worthwhile to emphasize that all results, for both BHM and WLS, were obtained with the very same set of sampled microstates. The only difference is on the way both methods evaluate $g(E)$. While the WLS measures $g(E)$ by changing an initial estimation for $g(E)$ by a controlled factor $f$ for every sampled microstate, the BHM, which is not bounded to any sampling dynamics, measures $g(E)$ with the macroscopic $N_{up,dn}^{\pm\Delta E}$ instead of simply incrementing the visits counter by unity. The benefit that our findings show is that without changing anything in the usual WLS simulation codes, we can have a more accurate result for the density of states. It only takes the extra work of evaluating the microcanonical averages $N_{up,dn}^{\pm\Delta E}$ during the usual WLS. 

Since the WLS is one of the most successful methods to estimate $g(E)$ and $\langle Q(E)\rangle$ (with flat histograms), we may conjecture that for any other microcanical dynamics one will attain a better accuracy for $g(E)$ with the BHM. Another advantage is that BHM can be applied to non-flat histograms of visits, since it is not based on this quantity. This way, one can taylor the optimum profile of visits along the energy axis \cite{BHM_BJP2, BJP_JSSM_PMCO}.  

\section{Acknowledgements}

This work was supported by the Brazilian agencies CNPq and CAPES. We thank Vitor Lara and Marlon Ramos for helpful discussions.

\bibliographystyle{apsrev4-1}
\bibliography{referencias_1.bib}

\end{document}